\documentclass[twocolumn,twoside,slac]{revtex4}
\usepackage{graphicx}
\usepackage{fancyhdr}
\usepackage[figuresright]{rotating}
\begin{document}

\title{Software Scalability Issues in Large Clusters}

%

\author{A. Chan, R. Hogue, C. Hollowell, O. Rind, T. Throwe, T. Wlodek}
\affiliation{Brookhaven National Laboratory, NY 11973, USA}

\begin{abstract}
The rapid development of large clusters built with commodity
hardware has highlighted scalability issues with deploying
and effectively running system software in large clusters.
We describe here our experiences with monitoring, image 
distribution, batch and other system administration software
tools within the 1000+ node Linux cluster currently in 
production at the RHIC Computing Facility.
\end{abstract}

\maketitle

\thispagestyle{fancy}

\section{BACKGROUND}

The rapid development of large clusters built with affordable
commodity hardware has highlighted the need for scalable cluster 
administration software that will help system administrators to
deploy, maintain and run large clusters. Scalable cluster 
administration software is critical for the efficient operation
of the 2000+ CPU Linux Farm at the RHIC Computing Facility (RCF),
and this paper describes our experience with cluster administration 
software currently in use at the RCF.

The RCF is a large scale data processing facility at Brookhaven 
National Laboratory (BNL) for the Relativistic Heavy Ion Collider 
(RHIC), a collider dedicated to high-energy nuclear physics 
experiments. The RCF provides for the computational needs of the RHIC 
experiments (BRAHMS, PHENIX, PHOBOS, PP2PP and STAR), including batch, 
mail, printing and data storage. In addition, BNL is the U.S. Tier 1 
Center for ATLAS computing, and the RCF also provides for the 
computational needs of the U.S. collaborators in ATLAS. 

The Linux Farm at the RCF provides the majority of the CPU power 
in the RCF. It is currently listed as the $3^{rd}$ largest cluster, 
according to "Clusters Top500" (http://clusters.top500.org). Figure 1 
shows the rapid growth of the Linux Farm in the last few years. 

All aspects of its development (hardware and software), operations
and maintenance are overseen by the Linux Farm group, currently a 
staff of 5 FTE within the RCF. 

\section{HARDWARE}

The Linux Farm is built with commercially available thin rack-mounted,
Intel-based servers (1-U and 2-U form factors). Currently, there are 
1097 dual-CPU production servers with approximately 917,728 SpecInt2000.
Table 1 summarizes the hardware currently in service in the Linux Farm.
Hardware reliability has not been an issue at the RCF. The average 
failure rate is 0.0052 $failures/(machine \cdot month)$, which translates
to 5.7 hardware failures per month at its present size. Hardware failures
are dominated by disk and power supply failures. A detailed breakdown of
the hardware failures by category is shown in Figure 2. 

\begin{table}[t]
\begin{center}
\caption{Linux Farm hardware}
\begin{tabular}{|c|c|c|c|c|}
\hline \textbf{Brand} & \textbf{CPU} & \textbf{RAM} &
\textbf{Storage} & \textbf{Quantity}
\\
\hline VA Linux & 450 MHz & 0.5-1 GB & 9-120 GB & 154\\
\hline VA Linux & 700 MHz & 0.5 GB & 9-36 GB & 48\\
\hline VA Linux & 800 MHz & 0.5-1 GB & 18-480 GB & 168\\
\hline IBM & 1.0 GHz & 0.5-1 GB & 18-144 GB & 315\\
\hline IBM & 1.4 GHz & 1 GB & 36-144 GB & 160\\
\hline IBM & 2.4 GHz & 1 GB & 240 GB & 252\\
 \hline
\end{tabular}
\end{center}
\end{table}

\section{MONITORING SOFTWARE}

Monitoring and control of the cluster hardware, software and 
infrastructure (power and cooling) is provided via a mix of 
open-source software, RCF-designed software and vendor-provided
software. Various components of the monitoring software have 
been redesigned for scalability purposes and to add various
persistent and fault-tolerant features to provide near real-time
information reliably. Figure 3 shows the two monitoring models
used in the design of the various components of the monitoring
software.  

Figure 4 shows the auto-updating Web-interface to the RCF-designed 
monitoring software, which allows us to view individual server 
status within the cluster. Figure 5 is the Web-interface to the
complementary open-source ganglia [1] monitoring software, which 
provides monitoring information in a user-friendly format via a Web 
interface.

\section{IMAGE DISTRIBUTION SOFTWARE}

The RCF used a Linux image distribution system that
relied on an image server that made the image available 
via NFS to the Linux Farm servers. The image was 
downloaded and deployed during the rebuilding process. 
That system worked well until the Linux Farm grew 
beyond $\sim 150$ servers, at which point NFS 
limitations prevented this system from reliably 
upgrading a large number of servers simultaneously in 

\begin{figure*}[t]
\begin{center}
\includegraphics[width=120mm]{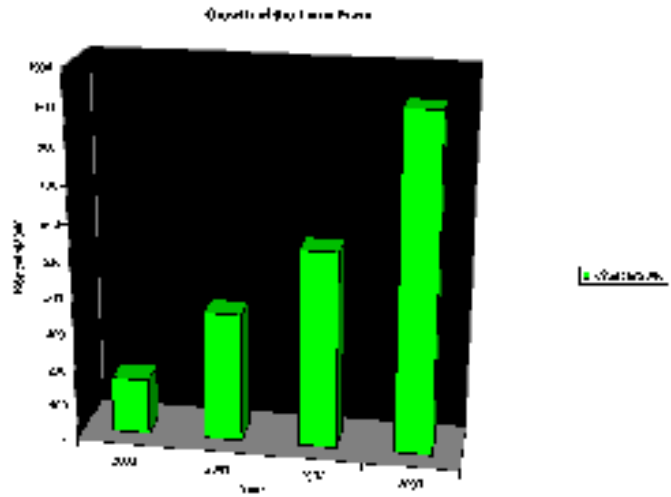}
\end{center}
\caption{The growth of the Linux Farm at the RCF.}
\end{figure*}
 
\begin{figure*}[b]
\begin{center}
\includegraphics[width=120mm]{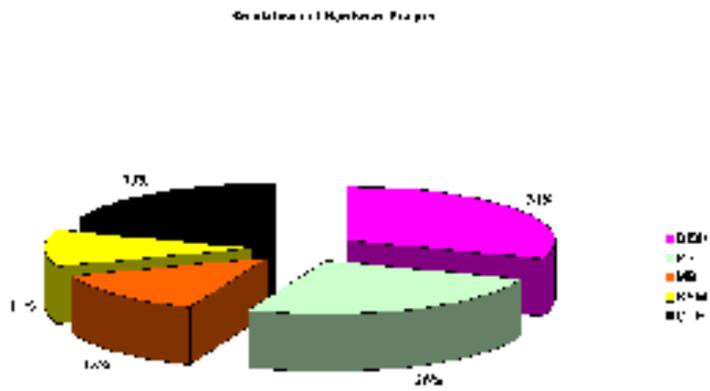}
\end{center}
\caption{Breakdown of hardware failure by category.}
\end{figure*}

\begin{figure*}[t]
\begin{center}
\includegraphics[width=160mm]{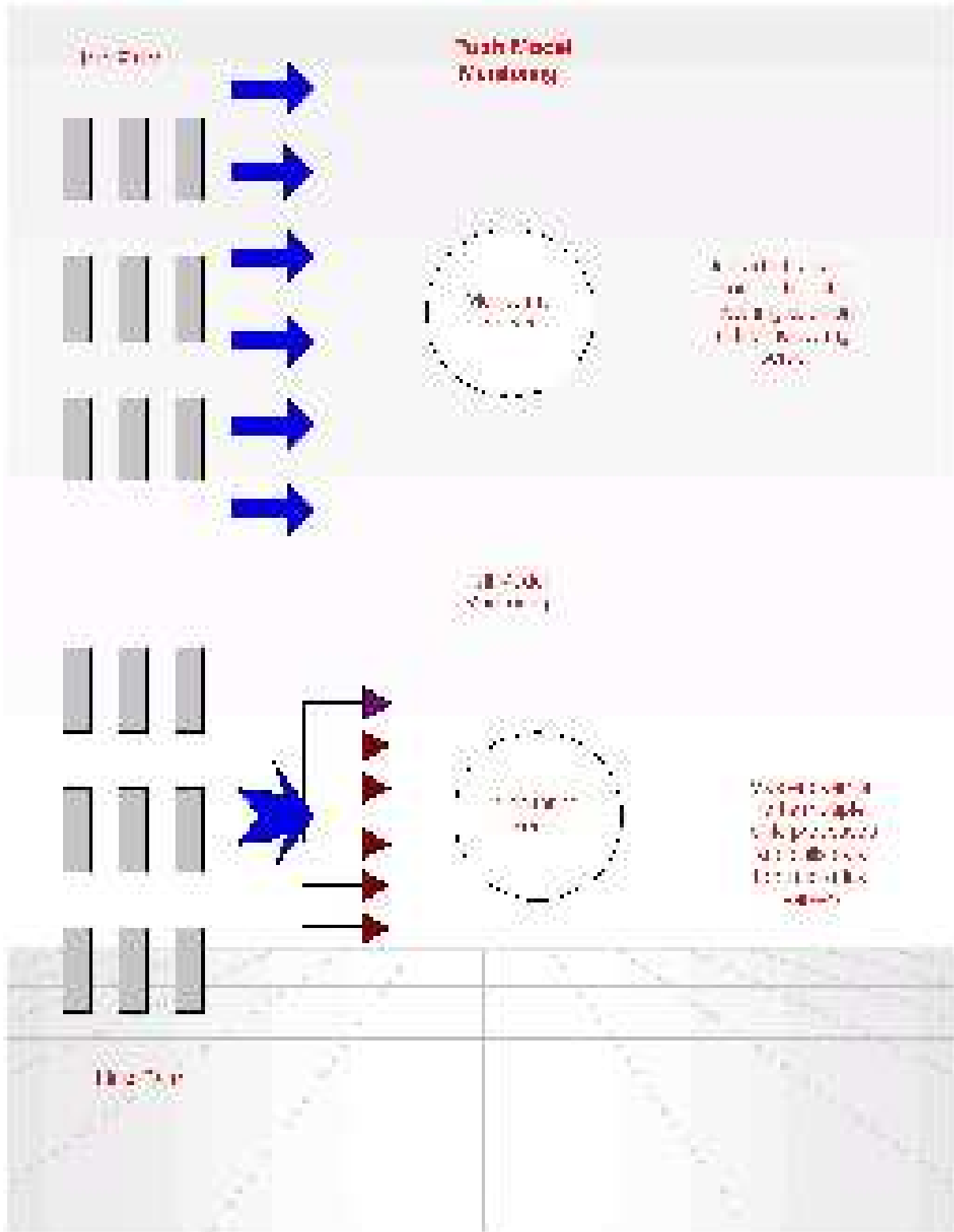}
\caption{Push vs. pull model for the RCF monitoring software.}
\end{center}
\end{figure*}

\begin{figure*}[t]
\begin{center}
\includegraphics[width=120mm]{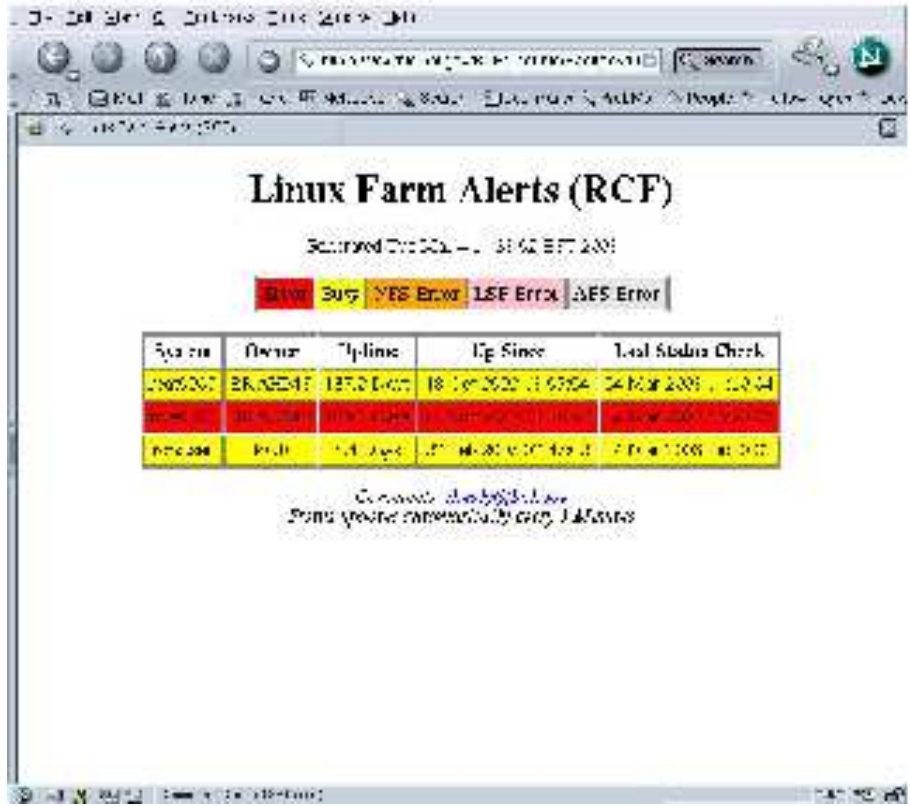}
\caption{System status of Linux Farm nodes.}
\end{center}
\end{figure*}

\begin{figure*}[b]
\begin{center}
\includegraphics[width=120mm]{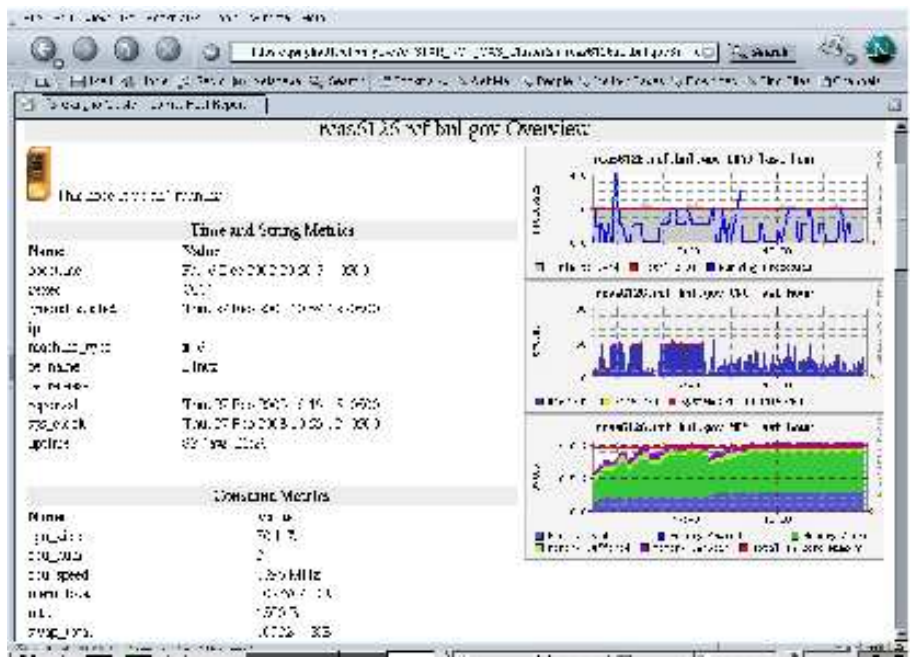}
\caption{Detailed view of a STAR node with ganglia.}
\end{center}
\end{figure*}

\begin{figure*}[t]
\begin{center}
\includegraphics[width=160mm]{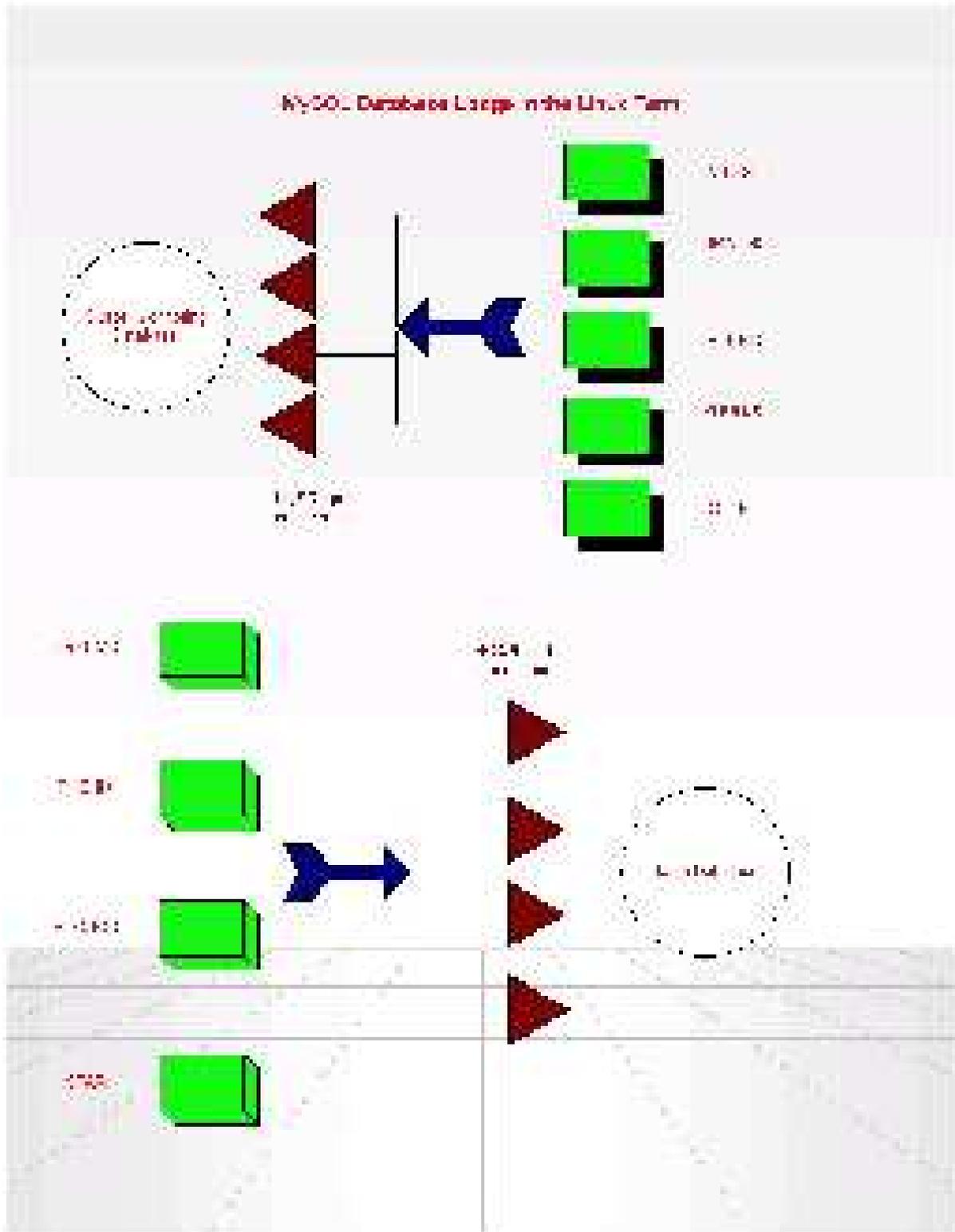}
\caption{MySQL usage in the RCF Linux Farm.}
\end{center}
\end{figure*}

\clearpage

\noindent
an acceptable time interval. 

In 2001, the RCF switched to RedHat's Kickstart installer 
[2]. Kickstart allowed us to switch to a Web-based image
server using standard rpm packages. It has proven very
scalable (20 minutes/server with 100's of servers rebuilt
simultaneously) and highly flexible. Multiple images can
co-exist with different install options. 

Both the old and the new Kickstart image distribution
systems rely on a secure MySQL [3] database system for 
server authentication and configuration specification.

\section{DATABASE SYSTEMS}

The RCF uses the open-source, lightweight MySQL database
software throughout the facility. MySQL enjoys wide 
support in the open-source community, and it is well 
documented. It is used as the general back-end data 
repository for monitoring, batch control and cluster 
management purposes at the RCF. MySQL is a scalable and
easily configurable database software. Figure 6 shows
how MySQL databases are configured in the Linux Farm to
achieve high scalability and reliability for monitoring
and batch control purposes. Figure 7 shows the PERL-TK
interface to the MySQL database for our batch control
and monitoring software. 

\begin{figure}[h]
\begin{center}
\includegraphics[width=80mm]{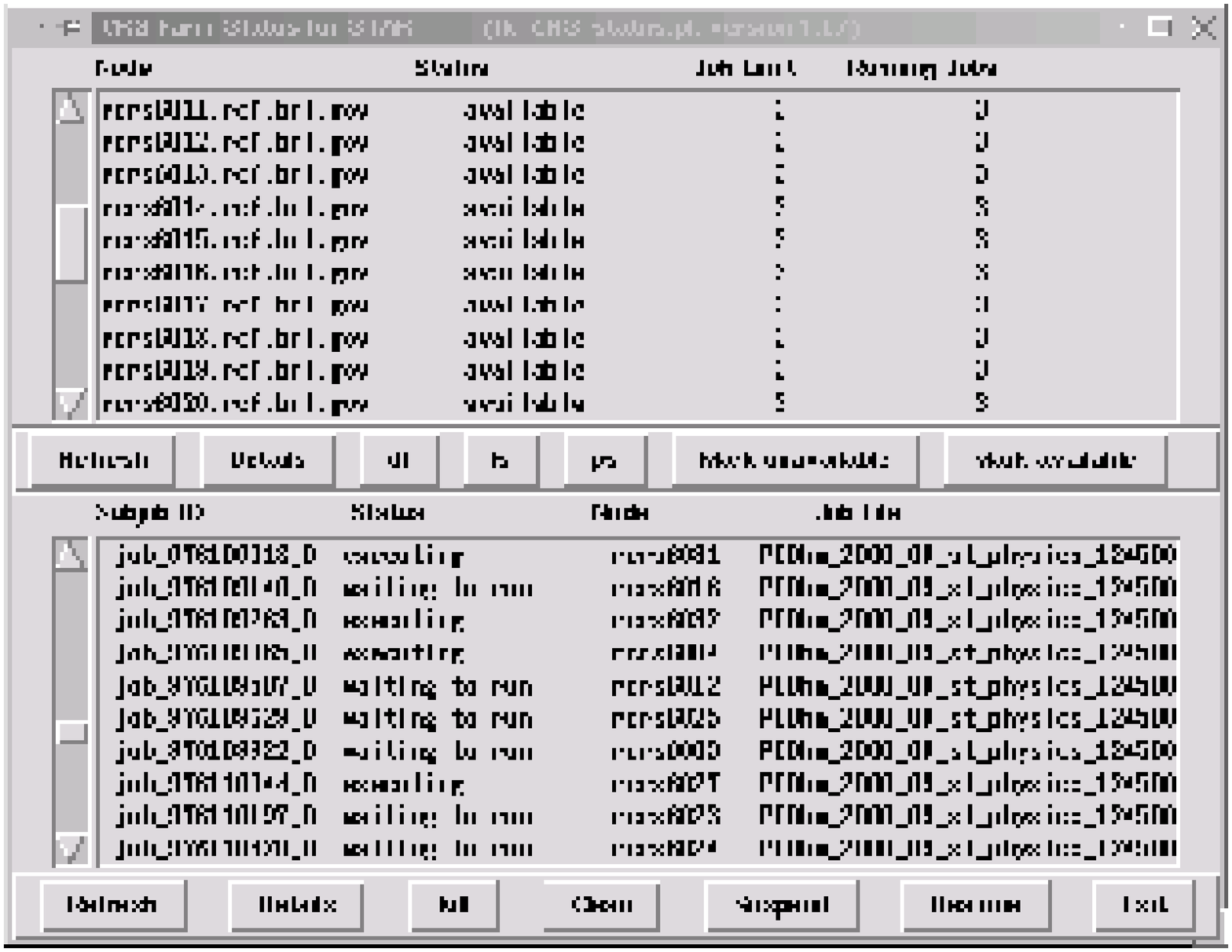}
\caption{Batch job control with MySQL back-end.}
\end{center}
\end{figure}

\subsection{OTHER SYSTEM ADMINISTRATION TOOLS}

The Linux Farm also uses RCF-designed PYTHON-based scripts
for fast, parallel access to the Linux Farm servers. The
scripts are used for routine system administration purposes
such as installing or updating software. The scripts are 
also used for automatic emergency remote power access to 
the servers during infrastrucure system failures (UPS or 
cooling). Because of the possibility of catastrophic
disasters in the case of infrastructure failures (for example,
an electrical fire), it is important that the scripts 
perform fast, parallel and controlled shutdown of the
Linux Farm servers. Figure 9 shows how our PYTHON-based 
scripts fork multiple processes to become a scalable 
system administration tool. 

The RCF also uses vendor-provided software for cluster
management. Figure 8 is an example of the user-interface
to a software system that is currently deployed in the RCF 
Linux Farm. 

\begin{figure}[h]
\begin{center}
\includegraphics[width=80mm]{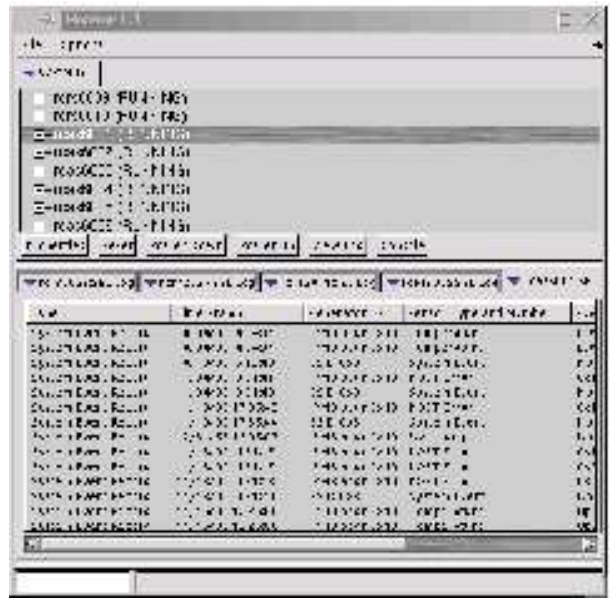}
\caption{Vendor provided remote power management software.}
\end{center}
\end{figure}

\section{CONCLUSION}

Scalable system software has become an important factor to
the RCF for efficiently deploying and managing our rapidly
growing Linux cluster. It allows us to monitor the status
of individual cluster servers in near-real time, to deploy
our Linux image in a fast and reliable fashion across the
cluster and to access the cluster in a fast, parallel manner.

Because not all of our system software needs can be addressed
from a single source, it has become necessary for us to use
a mix of RCF-designed, open-source and vendor-provided software
to achieve our goal of a scalable system software architecture.

\begin{figure*}[h]
\begin{center}
\includegraphics[width=160mm]{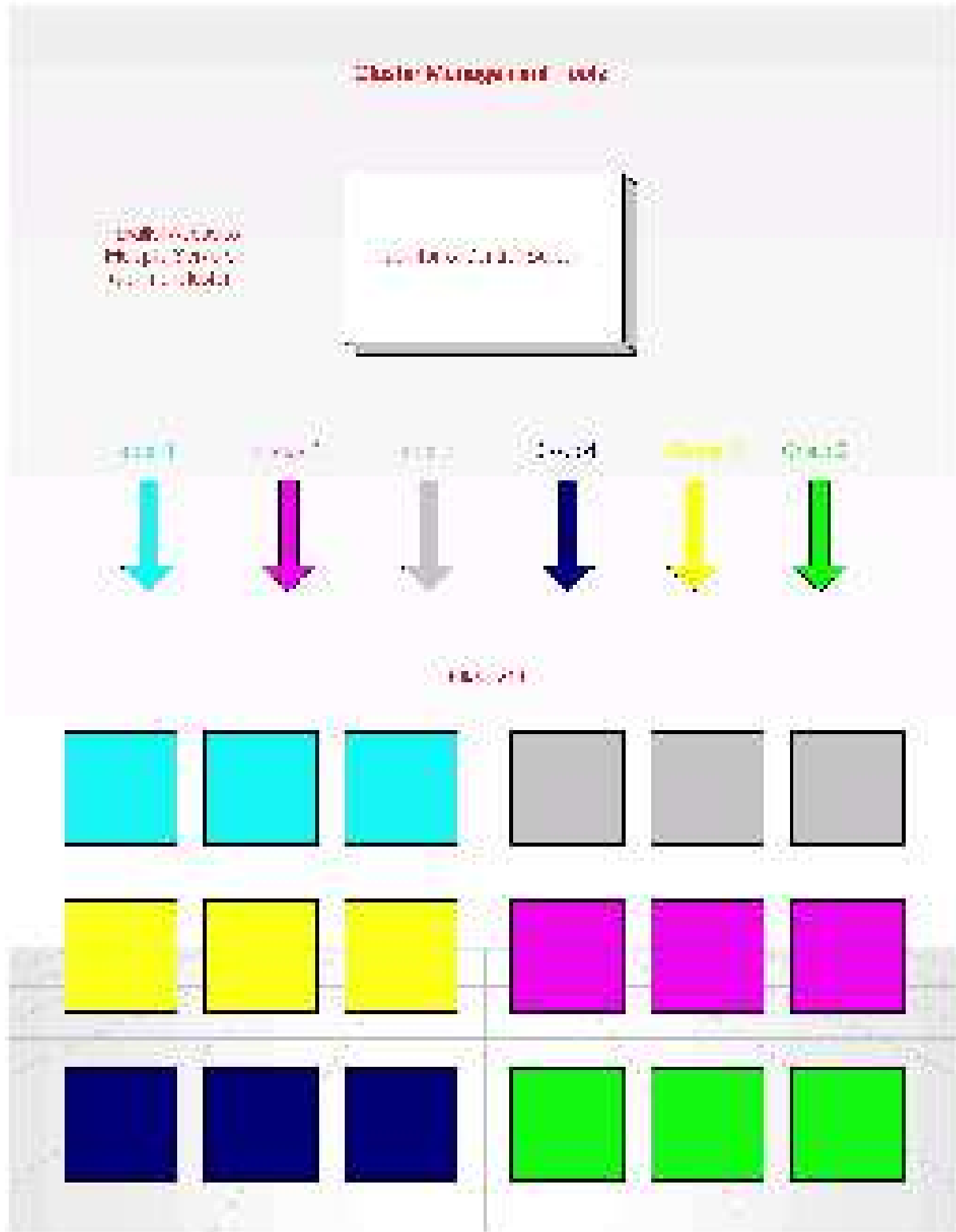}
\caption{Example of scalable cluster management tool.}
\end{center}
\end{figure*}

\clearpage

\begin{acknowledgments}
The authors wish to thank the Information Technology Division and
the Physics Department at BNL for their support to the RCF mission.

\end{acknowledgments}


\end{document}